\documentclass[%
 reprint,
 amsmath,amssymb,
 aps,
]{revtex4-2}

\usepackage{graphicx}% Include figure files
\usepackage{dcolumn}% Align table columns on decimal point
\usepackage{bm}% bold math
\usepackage{hyperref}% add hypertext capabilities
\usepackage[mathlines]{lineno}% Enable numbering of text and display math
\usepackage{longtable}
\usepackage{rotating}
\usepackage{multirow}
\usepackage{xcolor}
\newcommand{\jpsi}{J/\psi}
\newcommand{\psip}{\psi(2S)}

\newcommand{\elp}{e^{+}}
\newcommand{\elm}{e^{-}}
\newcommand{\mup}{\mu^{+}}
\newcommand{\mum}{\mu^{-}}
\newcommand{\pip}{\pi^{+}}
\newcommand{\pim}{\pi^{-}}

\newcommand{\prp}{p}
\newcommand{\prm}{\bar{p}}
\newcommand{\kp}{K^{+}}
\newcommand{\km}{K^{-}}
\newcommand{\ks}{K^{0}_{S}}
\newcommand{\kl}{K^{0}_{L}}

\newcommand{\Xf}{X_{f}}

\newcommand{\BESIIIorcid}[1]{\href{https://orcid.org/#1}{\hspace*{0.1em}\raisebox{-0.45ex}{\includegraphics[width=1em]{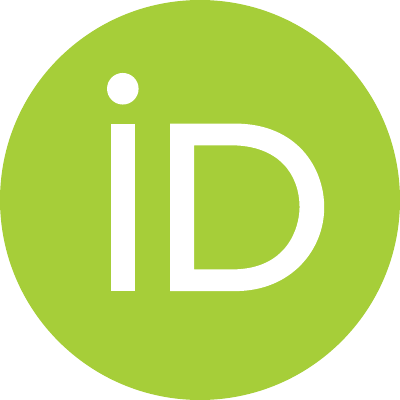}}}}

\begin{document}
%\linenumbers
\preprint{APS/123-QED}
\bibliographystyle{apsrev4-1}

\title{First energy scan measurement of $\elp\elm\to\kp\km$ around the $\psip$ resonance}
\author{
\begin{small}
\begin{center}
%% Saved at => 2025-08-15
M.~Ablikim$^{1}$\BESIIIorcid{0000-0002-3935-619X},
M.~N.~Achasov$^{4,b}$\BESIIIorcid{0000-0002-9400-8622},
P.~Adlarson$^{81}$\BESIIIorcid{0000-0001-6280-3851},
X.~C.~Ai$^{86}$\BESIIIorcid{0000-0003-3856-2415},
R.~Aliberti$^{39}$\BESIIIorcid{0000-0003-3500-4012},
A.~Amoroso$^{80A,80C}$\BESIIIorcid{0000-0002-3095-8610},
Q.~An$^{77,64,\dagger}$,
Y.~Bai$^{62}$\BESIIIorcid{0000-0001-6593-5665},
O.~Bakina$^{40}$\BESIIIorcid{0009-0005-0719-7461},
Y.~Ban$^{50,g}$\BESIIIorcid{0000-0002-1912-0374}
% ...
\\
\vspace{0.2cm}
(BESIII Collaboration)\\
\vspace{0.2cm} {\it
$^{1}$ Institute of High Energy Physics, Beijing 100049, People's Republic of China\\
% ...
$^{4}$ Budker Institute of Nuclear Physics SB RAS (BINP), Novosibirsk 630090, Russia\\
% ...
$^{39}$ Johannes Gutenberg University of Mainz, Johann-Joachim-Becher-Weg 45, D-55099 Mainz, Germany\\
$^{40}$ Joint Institute for Nuclear Research, 141980 Dubna, Moscow region, Russia\\
% ...
$^{50}$ Peking University, Beijing 100871, People's Republic of China\\
% ...
$^{62}$ Southeast University, Nanjing 211100, People's Republic of China\\
% ...
$^{64}$ State Key Laboratory of Particle Detection and Electronics, Beijing 100049, Hefei 230026, People's Republic of China\\
% ...
$^{77}$ University of Science and Technology of China, Hefei 230026, People's Republic of China\\
% ...
$^{80}$ University of Turin and INFN, (A)University of Turin, I-10125, Turin, Italy; (B)University of Eastern Piedmont, I-15121, Alessandria, Italy; (C)INFN, I-10125, Turin, Italy\\
$^{81}$ Uppsala University, Box 516, SE-75120 Uppsala, Sweden\\
% ...
$^{86}$ Zhengzhou University, Zhengzhou 450001, People's Republic of China\\

\vspace{0.2cm}
$^{\dagger}$ Deceased\\
% ...
$^{b}$ Also at the Novosibirsk State University, Novosibirsk, 630090, Russia\\
% ...
$^{g}$ Also at State Key Laboratory of Nuclear Physics and Technology, Peking University, Beijing 100871, People's Republic of China\\
% ...

}
%% ends here %%
 
\end{center}
\end{small}
}

\collaboration{BESIII Collaboration}%\noaffiliation

\date{\today}% It is always \today, today,

\begin{abstract}
We report the precise energy scan measurement of the cross section lineshape of $\psip\to\kp\km$. The analysis is based on $\elp\elm$ collision data corresponding to an integrated luminosity of 495~pb$^{-1}$ collected with the BESIII detector at BEPCII. By analyzing the cross section line-shape, we extract the relative phase $\Phi$ between the strong and electromagnetic amplitudes of the $\psip$ resonance, a fundamental parameter in charmonium physics, based on the assumption that the relative phase between the electromagnetic amplitude of the $\psip$ resonance and the continuum is zero. Two distinct solutions for the branching fraction $\mathcal{B}$ of $\psip\to\kp\km$ are observed: a constructive interference solution with $\mathcal{B}=(7.49\pm0.41)\times10^{-5}$ and $\Phi=(110.1 \pm6.7)^\circ$, and a destructive interference solution with $\mathcal{B}=(10.94\pm0.48)\times10^{-5}$ and $\Phi=(-106.8\pm5.7)^\circ$. A significant correlation between $\Phi$ and $\mathcal{B}$ is established, demonstrating that interference effects must be taken into account in the $\psip$ branching fraction measurements. Additionally, the first results for both the $\psip$ strong form factor, which characterizes the strong coupling between $\psip$ and $\kp\km$, and the energy-dependent electromagnetic form factor of the charged kaon in this energy region are reported. 
\end{abstract}

\maketitle

%\tableofcontents
The relative phase $\Phi$ between the strong and electromagnetic (EM) amplitudes of the $J^{\rm PC}=1^{--}$ states is a fundamental parameter in particle physics, serving as a crucial input for studying quarkonium decays, particularly in charmonium systems.
The Feynman amplitude for vector charmonium states below the $D\bar{D}$ threshold decaying into hadrons consists of two components: a three-gluon mediated strong amplitude $A_g$ and a virtual-photon mediated EM amplitude $A_\gamma$. Studies~\cite{Suzuki:1998ea,Suzuki:1999nb,Suzuki:2000yq,Rosner,Baldini:1998en,Baldini:1996hc} reveal a striking relative phase $\Phi$ of $\pm90^\circ$ between them in $\jpsi$ decays, strongly indicating that the $\Phi$ could have a universal meaning~\cite{Gerard,Suzuki:1998ea} in different decay modes. 
On the other hand, the analyses based on SU(3) flavor symmetry~\cite{Haber:1985cv} of the measured branching fractions, most of which were obtained over 30 years ago, reported very different values in different charmonium decays~\cite{Suzuki:2000yq,Wang:2012mf,Metreveli,Mo:2024jjd}. 
% check Rosner's paper about the phase theory, Moxiaohu's paper has mentioned some.
These discrepancies in the phase measurement are used to address the $\rho\pi$ and non-$D\bar{D}$ decay puzzles~\cite{Li:2008fk,Zhao:2010zzv,Kivel:2023fgu,Wang:2002np}. However, Refs.~\cite{LopezCastro:1994xw,Gerard} discuss that the $\Phi$ should not differ significantly among charmonia because the three-gluon amplitude primarily involves the charm quark mass scale. The origin of this possible universal phase and its explanation remain unclear.

The decays of charmonia into pseudoscalar-anti-pseudoscalar meson pairs, $P\bar{P}$, are ideal platforms to test whether the phase is universal through SU(3) symmetry analysis. 
Within this model, the charmonium decaying into $\pip\pim$ is purely electromagnetic with $A_{g}$ forbidden by isospin invariance. The $\ks\kl$ decay is purely strong in the limit of SU(3) symmetry, while the $\kp\km$ decay can proceed through both $A_{g}$ and $A_{\gamma}$. 
%Thus, charmonium decaying into pseudoscalar pairs (PP) provides one of the best places for studying the phase difference.
Furthermore, according to the 12\% rule~\cite{Appelquist}, the amplitude $A_{g}$ of $\psip\to\kp\km$ is weaker than that of $J/\psi \to K^+K^-$, thereby enhancing the visibility of strong-EM interference effects compared to those in the $\jpsi$ decay. 
An SU(3) analysis~\cite{Yuan} based on the branching fractions of the decays $\psip\to K^0_S K^0_L,K^+K^-,\pi^+\pi^-$ measured by BES~\cite{Bai} gave $\Phi=(-82\pm29)^\circ$ or $\Phi=(121\pm27)^\circ$. 
The results were subsequently confirmed by two independent analyses~\cite{Metreveli,Metreveli0} using CLEO data, which moreover observed a discrepancy of $(37.0^{+16.5}_{-10.5})^\circ$ between $\psip$ and $\jpsi$. 
Furthermore, the BaBar experiment proved that the interference cannot be ignored, especially for charmonium decays, by verifying that the interference shifts the measured branching fractions of $\pm5\%$ for the $\jpsi$ meson and of $\pm15\%$ for the $\psip$ meson~\cite{ref_exp_KFF2}. 

Around the $\psip$, the $e^+e^-\to\kp\km$ process receives both photon-mediated continuum amplitude $A_{\rm cont}$ and charmonium-mediated contributions. Measurements of the $\elp\elm\to\mup\mum$ cross section around the $\jpsi$ resonance show that the amplitudes $A_{\rm cont}$ and $A_\gamma$ have a relative phase $\Phi_{\gamma,\rm cont}$ compatible with zero~\cite{KEDR,J5pi}. By assuming $\Phi_{\gamma,\rm cont}$ is universal, the parameter $\Phi$ can be extracted by analyzing the line shape of the measured hadronic cross sections around the resonance.

In addition, the measured cross sections for $e^+e^-\to K^+K^-$  around the $\psip$ also allow to extract the EM energy-dependent charged kaon form factor and the constant strong one. 
The adjective ``strong'' emphasizes the role of the coupling constant between the $\psi(2S)$ resonance and the $\kp\km$ final state.
The most general parameterization of the transition matrix $\mathcal{M}^\mu_{\rm EM}$~\cite{Rosini:2025wfm} of the process $\gamma^*\to \kp\km$, with charged kaon being a pseudoscalar meson, contains only one EM form factor, $F_K(s)$, {\it i.e.}, 
\begin{equation}
\mathcal{M}^\mu_{\rm EM} = e F_K(s) \left(p_+^\mu-p_{-}^\mu\right) \notag \,, 
\end{equation}
where $e$ is the positron electric charge, $p_{+(-)}$ is the four-momentum of $K^{+(-)}$ and $s=\left(p_++p_{-}\right)^2$ is the squared four-momentum of the virtual photon. Analogously, the transition matrix of the process $\psi(2S)\to 3g\to \kp\km$ is
\begin{equation}
\mathcal{M}^\mu_{g} = e f_K \left(p_+^\mu-p_{-}^\mu\right) \notag \,,  
\end{equation}
where $f_K$ is the strong form factor,
%, and $\psi(2S)$ stands for $\psip$.
%\textcolor{red}{The EM form factor is a constant}, 
and it is defined only at the $\psi(2S)$ mass: $s=\left(p_++p_-\right)^2=m_{\psi(2S)}^2$. %The adjective ``strong'' emphasizes the role of coupling constant between the $\psi(2S)$ resonance and the $\kp\km$ final state by $f_K$.

The amplitudes of the processes $e^+e^-\to\gamma^*\to \kp\km$, $e^+e^-\to\gamma^*\to\psi(2S)\to\gamma^*\to\kp\km$, $e^+e^-\to\gamma^*\to\psi(2S)\to 3g\to \kp\km$,  are given by 
\begin{eqnarray}
&A_{\rm cont}&=e J_\mu^e \frac{-i}{s}F_K(s) \left(p_+^\mu-p_{-}^\mu\right)\,,\notag \\
&A_{\gamma}&=e J_\mu^e \frac{-i}{s}\frac{-i(i g_{\psi(2S)}^\gamma)^2(s/M^2)}{s-M^2+iM\Gamma} \frac{-i}{s} F_K(s) \left(p_+^\mu-p_{-}^\mu\right)\,, \notag \\
&A_{g}&=e J_\mu^e \frac{-i}{s} \frac{-i(ig_{\psi(2S)}^\gamma) (s/M^2)}{s-M^2+iM\Gamma} f_K \left(p_+^\mu-p_{-}^\mu\right) \notag \,, 
\end{eqnarray}
where $J^e_\mu$ is the leptonic current of the common $e^+e^-$ initial state, $-i/s$ is the photon propagator, $ig_{\psi(2S)}^\gamma$ is the coupling constant of virtual photon and $\psi(2S)$ charmonium, $-i(s/M^2)/\left(s-M^2+iM\Gamma\right)$ is the Breit-Wigner propagator of the charmonium $\psi(2S)$ with mass $M$ and width $\Gamma$. The term $(s/M^2)$ ensures the vanishing of the contribution at $s= 0$ and equals to $1$ at the $\psi(2S)$ mass. In the $A_\gamma$ amplitude, the $(ig^\gamma_{\psi(2S)})^2$ term is due to an $ig_{\psi(2S)}^\gamma$ for the first conversion $\gamma^*\to\psi(2S)$ and another for the subsequent re-conversion $\psi(2S)\to\gamma^*$. This coupling constant has the dimension of energy squared and is proportional to the square root of the electronic width $\Gamma^{0}_{ee}$ of the decay $\psi(2S)\to e^+e^-$, and it is
\begin{eqnarray}
g_{\psi(2S)}^\gamma=\sqrt{{3\,\Gamma^{0}_{ee} M^3}/{\alpha}} \notag \,,    
\label{eq:g-gamma-psi}
\end{eqnarray}
where $\alpha=1/137$ is the fine structure constant. The total amplitude is defined as 
\begin{eqnarray}
    &&A=A_{\rm cont}+A_{\gamma}+A_g
    =e J_\mu^e\frac{-i}{s}F_K(s)\left(p_+^\mu-p_{-}^\mu\right)\cdot
\notag   \\
  &&\left[1+\left(1+\frac{f_K}{F_K(s)}\frac{M^2}{g_{\psi(2S)}^\gamma}\right)\frac{(g_{\psi(2S)}^\gamma)^2(s/M^2)/M^2}{s-M^2+iM\Gamma}
    \right]\,,
    \label{eq:a-tot}
\end{eqnarray}
where the second photon propagator of $A_\gamma$ has been fixed at the $\psi(2S)$ mass. The value of $\Phi$ is expressed as
\begin{eqnarray}
    \Phi={\rm arg}\left({f_K}/{F_K(M^2)}\right) \notag \,. 
 \end{eqnarray}

In this Letter, we report the first scan measurement of the cross section lineshape of $\psip\to\kp\km$ using $\elp\elm$ collision data collected at nine center-of-mass (c.m.) energies from 3.58~GeV to 3.71~GeV corresponding to a total integrated luminosity of $495$~pb$^{-1}$. 
By analyzing the cross section line-shape and independently from SU(3), the first and direct results of the phase between strong and EM amplitudes and the corresponding branching fractions in different interference scenarios are extracted for $\psip\to\kp\km$, as well as the EM and the strong form factors of the charged kaon. 
The integrated luminosities of each datasets are measured with di-photon events to avoid interference between the resonance and continuum production, following Ref.~\cite{BESIII:2017lkp}.

The BESIII detector is described in detail in Ref.~\cite{BESIII}. Signal and background MC samples are generated using: ConExc~\cite{conexc} with vector-scalar-scalar modeling the angular distributions to match theoretical predictions; BabaYaga (version 3.5)~\cite{babayaga} for Bhabha, $\mup\mum$, and $\pip\pim$ backgrounds; and LUARLW~\cite{lund} for inclusive hadronic events. All simulations incorporate full GEANT4 detector modeling via BOSS~\cite{BESIII,Geant4} for efficiency and background studies.

Candidate events are required to have two charged particles, which are detected in the multilayer drift chamber (MDC) and within a $\theta$ polar angle range of $|\cos\theta|<0.93$, where $\theta$ is defined with respect to the $z$-axis, the symmetry axis of the MDC. For charged tracks, the distance of the closest approach to the interaction point (IP) must be less than 10~cm along the symmetry axis of the MDC, and less than 1~cm in the transverse plane. Particle identification (PID) is applied where the specific ionization energy loss $\mathrm{d}E / \mathrm{d}x$, measured by the MDC, and the flight time, measured using the time-of-flight (TOF), form likelihoods $\mathcal{L}$ for $K$ or $\pi$ hypothesis. Both tracks are required to be identified as kaons with $\mathcal{L}(K)>\mathcal{L}(\pi)$.

To suppress Bhabha events, the requirement $E/p<0.8 c$ is imposed for the charged tracks, representing the ratio between the energy deposited in the EM calorimeter and the momentum measured in the MDC. Cosmic rays are rejected by requiring $\Delta T\equiv |T_{\kp}-T_{\km}|<3$~ns, where $T_{\kp}$ and $T_{\km}$ are the measured flight time from the interaction point to the TOF detector for the two tracks. 
To reduce backgrounds from events with additional tracks, the residual momentum $\Delta P=|P_{\rm c.m.}-P_{\kp}-P_{\km}|$ is required to be less than 0.07~GeV$/c$. 
\vspace{-3mm}
\begin{figure}[htbp]
\includegraphics[angle=0,width=0.5\columnwidth]{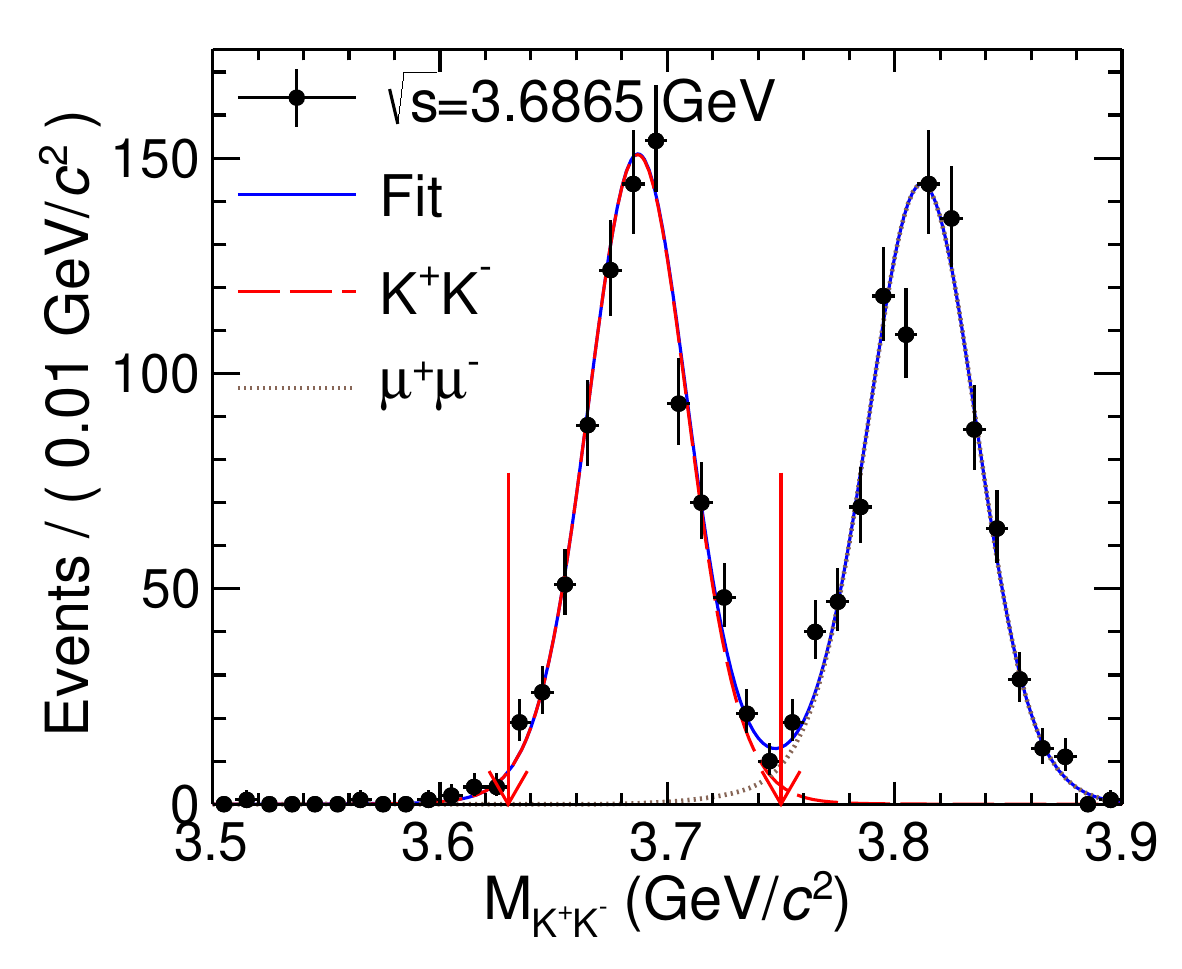}\put(-20,80){(a)} 
\includegraphics[angle=0,width=0.5\columnwidth]{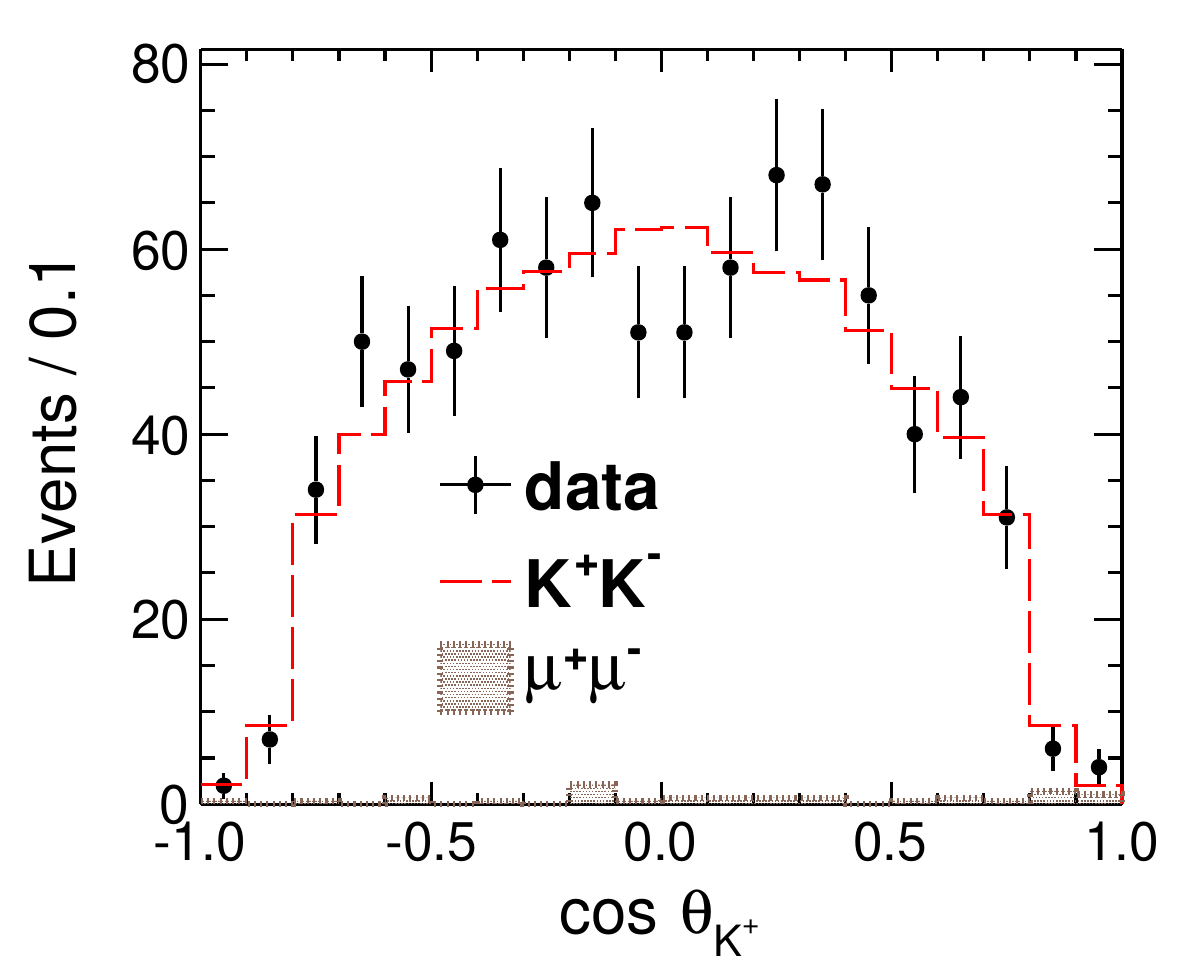}\put(-20,80){(b)}
\vspace{-3mm}
\caption{(a) Distribution of $M_{K^+K^-}$ of the accepted candidates at $\sqrt{s}=3.6865$~GeV. (b) Distribution of cosine of the polar angle $\theta$ of $\kp$ in the signal region for all energy points. The dots with error bars represent the data. The red dashed curve and histogram are the signal MC sample, and the brown dotted line and filled histogram are the background events of $\elp\elm\to\mup\mum$, respectively. The region between the arrows denotes the signal region.}
\label{fig_mkk}
\end{figure}

As an example, Fig.~\ref{fig_mkk}(a) shows the distribution of $\kp\km$ invariant mass, $M_{K^+K^-}$, at the c.m. energy $\sqrt{s}=3.6865$~GeV. The dominant background is from the process $\elp\elm\to\mup\mum$ that is described by MC simulation. The other backgrounds from two-body ($\prp\prm$ and $\pip\pim$) or multi-body hadronic decays are negligible from MC simulation estimation. 
The signal region is defined as the interval of 0.12~GeV around $\sqrt{s}$, which contains more than 99\% of signal events. For events selected in the signal region of all datasets, the angular distribution of $\kp$ in the c.m. frame is compared with the signal MC sample, as shown in Fig.~\ref{fig_mkk}(b). The consistency is noticeable. %, except for the accumulation of events around $\cos\theta_{\kp}=1.0$, which is due to the Bhabha background and would be excluded in the fit to the $M_{K^+K^-}$ distribution.  

\begin{figure}[htbp]
\begin{center}
\includegraphics[angle=0,width=1.0\columnwidth]{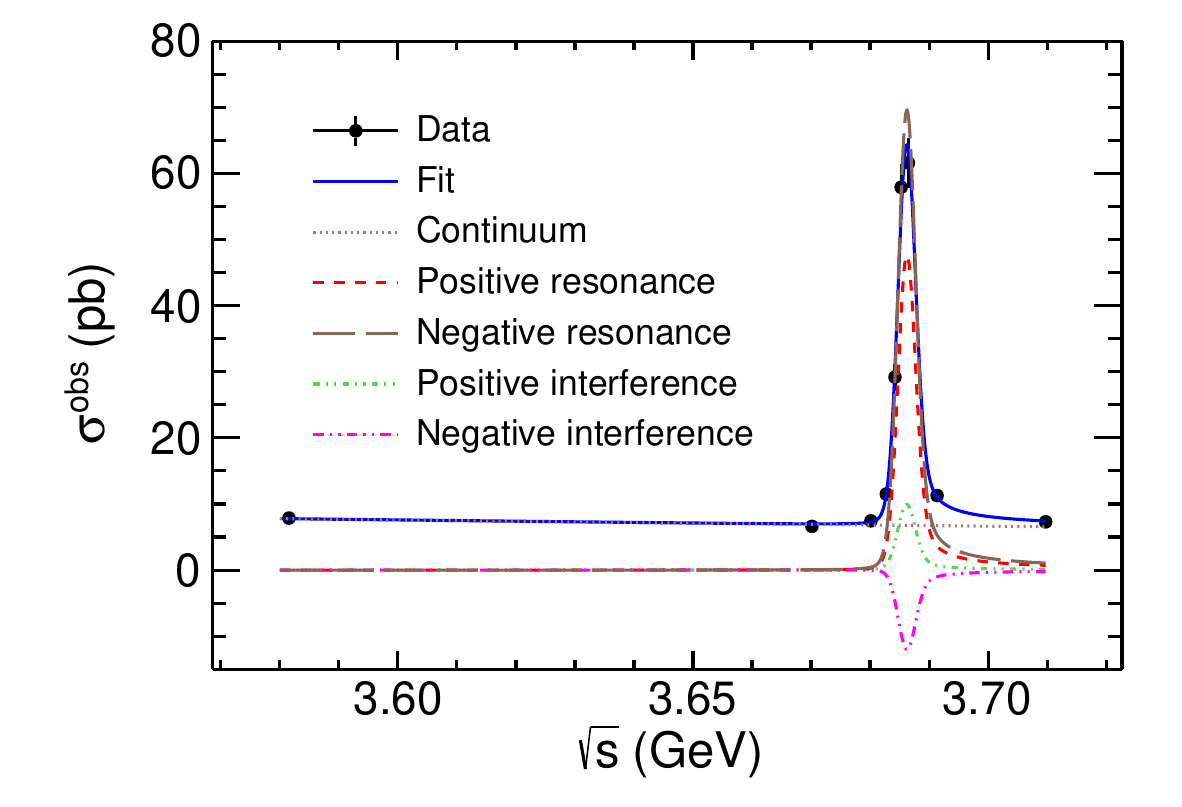}
\caption{The $\elp\elm\to\kp\km$ cross section data (black points) are shown with statistical and systematic uncertainties. The blue line is the fit result, with the dotted line indicating the continuum. The dashed and dash-dotted lines represent resonance and interference contributions, respectively.}
\label{fig_xs}
\end{center}
\end{figure}

An unbinned maximum likelihood fit is performed on the $M_{\kp\km}$ distribution to extract signal yield, as shown in Fig.~\ref{fig_mkk}(a) as an example. Two simulated shapes are used to describe the signal candidates and the background events from $\mup\mum$.
% or Bhabha. 
Each of them is convolved with a Gaussian function to account for the discrepancies in mass and resolution between data and MC. 
The observed cross section is determined as
\begin{equation}
\sigma^{\rm obs}={N^{\rm obs}}/({\epsilon\,\mathcal{L}}), \notag
\end{equation}
where $N^{\rm obs}$ is the number of signal events from the fit, $\mathcal{L}$ is the integrated luminosity and $\epsilon$ is the detection efficiency after iterations following the procedure implemented in Ref.~\cite{J5pi}. The obtained cross sections at all energy points are shown in Fig.~\ref{fig_xs}, and the values are listed in Table~\ref{tab_xs}. 
\begin{table*}
%\begin{sidewaystable}[h]
\caption{The integrated luminosity (${\cal L}$), observed signal yield ($N_{\rm obs}$), efficiency $\epsilon$, vacuum polarization factor $\frac{1}{|1-\Pi_0|^2}$, observed cross section $\sigma^{\rm obs}$ and absolute values of the charged kaon EM form factor $|F_{K}|$ at all energy points are shown. The uncertainties of $N_{\rm obs}$ and $\epsilon$ are statistical. 
The first uncertainties of cross sections are statistical, the second the systematic, and the third in the bracket are total uncorrelated systematic. %The uncertainty of $(1+\delta)$ is the difference between the latest two iterations.
}\label{tab_xs}
%\begin{center}
\begin{ruledtabular}
\begin{tabular}{ccccccc} 
$\sqrt{s}~{\rm (MeV)}$ &$\mathcal{L}$~(pb$^{-1}$) &$N_{\rm obs}$  &$\epsilon$~(\%)   &$\frac{1}{|1-\Pi_0|^2}$ &$\sigma^{\rm obs}$ ~(pb)    &$|F_{K}| (\times 100)$  \\ \hline
$3581.5\pm0.1$  &$84.604\pm0.082$    &$337.0\pm19.2$   &$50.6\pm0.1$  &$1.0355$ &$ 7.87\pm0.45\pm0.41(0.15)$  &  $7.28\pm 0.14$\\
$3670.2\pm0.1$  &$83.582\pm0.084$    &$273.4\pm17.3$   &$49.4\pm0.1$  &$1.0335$ &$ 6.62\pm0.42\pm0.43(0.29)$  &  $6.93\pm 0.13$\\
$3680.1\pm0.1$  &$83.060\pm0.083$    &$307.1\pm18.4$   &$49.6\pm0.1$  &$1.0333$ &$ 7.45\pm0.45\pm0.37(0.09)$  &  $6.90\pm 0.13$\\
$3682.8\pm0.1$  &$28.175\pm0.049$    &$167.0\pm13.5$   &$52.4\pm0.1$  &$1.0333$ &$11.51\pm0.92\pm0.59(0.18)$  &  $6.89\pm 0.13$\\
$3684.2\pm0.1$  &$27.840\pm0.048$    &$451.7\pm21.7$   &$55.6\pm0.1$  &$1.0333$ &$29.19\pm1.40\pm1.50(0.49)$  &  $6.88\pm 0.13$\\
$3685.3\pm0.1$  &$25.342\pm0.046$    &$827.6\pm29.2$   &$56.4\pm0.1$  &$1.0332$ &$57.89\pm2.04\pm2.95(0.87)$  &  $6.88\pm 0.13$\\
$3686.5\pm0.1$  &$24.481\pm0.045$    &$852.7\pm29.5$   &$56.6\pm0.1$  &$1.0332$ &$61.58\pm2.13\pm3.07(0.67)$  &  $6.87\pm 0.13$\\
$3691.4\pm0.1$  &$68.647\pm0.076$    &$400.3\pm20.6$   &$51.7\pm0.1$  &$1.0331$ &$11.28\pm0.58\pm0.57(0.14)$  &  $6.85\pm 0.13$\\
$3709.8\pm0.1$  &$69.326\pm0.077$    &$248.7\pm16.5$   &$49.1\pm0.1$  &$1.0326$ &$ 7.30\pm0.48\pm0.40(0.18)$  &  $6.79\pm 0.13$\\
\end{tabular}
\end{ruledtabular}
%\end{center}
\end{table*}
%\end{sidewaystable}

The systematic uncertainties of the cross section measurements include  correlated sources, related to the event selections, MC simulation and the luminosity measurement, as well as uncorrelated sources, which mainly arise from the fit to the $M_{K^+K^-}$ spectrum. 
In the line-shape analysis, the minimization is performed using a parameterized $\chi^2$ method where common systematic uncertainties are incorporated as the associated uncertainty of the scaling factor, while uncorrelated uncertainties are directly included in the $\chi^2$ calculation~\cite{J5pi}.

The systematic uncertainty of tracking is estimated to be 1.0\% per track with the control sample $\elp\elm\to K^{+}K^{-}\pi^{0}$.
The PID systematic uncertainty is determined by studying the $\kp$ and $\km$ efficiencies separately using the control sample, where the PID condition is not used on the track under study and the muon counter is used to enhance purity.
%The PID efficiency is studied with the $\psip\to\kp\km$ control sample, where the PID condition is not applied to the $\kp$ track and the muon counter information is used to improve the purity. The event yield is extracted from the $M_{\kp\km}$ spectrum. 
The difference in efficiency between data and MC simulation, $\sim2.0\%$ per track, is assigned as the systematic uncertainty. 
The uncertainty in the requirement of the residual momentum is estimated by fitting the distribution, and the difference in efficiency between data and MC simulation, 1.0\%, is taken as the systematic uncertainty. The uncertainty in the $\Delta T$ selection is estimated in a similar way and is found to be negligible. The uncertainty related to the $E/p$ requirement is estimated by varying the corresponding selection criteria, and a common value of $1.3\%$ is given from a fit on the largest difference in cross section at each energy point. The uncertainty due to the MC generator is assigned as the difference in obtained cross sections between {\sc conexc} and the line-shape fit program~\cite{ana_form}, which is 1.0\%.

The uncertainty in the integrated luminosity is $1.0\%$~\cite{BESIII:2017lkp}. To study the uncertainties resulting from the signal and background shapes, the fixed resolution of the Gaussian function is varied within $1\sigma$. Besides, the MC shapes for $\elp\elm\to\kp\km$ and $\elp\elm\to\mup\mum$ are alternatively described by two Crystal Ball functions. The fit range is varied by $\pm$0.1~GeV/$c^{2}$ to estimate its uncertainty. 
The total systematic uncertainty is determined by adding all these sources in quadrature, as listed in Table~\ref{tab_xs}. 

The description of the dressed cross section is obtained from the total amplitude of Eq.~\eqref{eq:a-tot}, as:
\begin{eqnarray}
\label{eq_Born}
\sigma^{0}(\sqrt{s})&=&\beta_{K}^3\left(\frac{\mathcal{F}}{s}\right)^2\frac{4\pi\alpha^{2}}{3s}\frac{1}{|1-\Pi_{0}(s)|^2} \notag \\
&&\cdot \left|1+(1+\mathcal{C}e^{i\Phi}) \frac{s}{M} \frac{3 \Gamma^{0}_{ee}/\alpha}{s-M^{2}+iM\Gamma}\right|^{2},
\end{eqnarray}
where $\Gamma_{ee}=\Gamma^{0}_{ee}/|1-\Pi_{0}(s)|$ is the measured electronic width of $\psip$ and cited from PDG as well as the values of $M$, $\Gamma$~\cite{pdg}. The vacuum polarization factor $1/|1-\Pi_{0}(s)|^2$ is calculated with the program developed in Ref.~\cite{Berends}. The parameter $\mathcal{C}\equiv|A_g|/|A_{\gamma}|$ is proportional to $|f_K/F_K(M^2)|$ at the $\psip$ mass (Eq.~\ref{eq:a-tot}), while $\beta_{K}=\sqrt{1-4m^2_K/s}$ is the kaon velocity. 
In particular, the kaon EM form factor \footnote{The factor of two comes from the comparison between the Born cross section of Eq.~\eqref{eq_Born} and 
\begin{eqnarray}  
\sigma^0(\sqrt{s})=\frac{\pi\alpha^2}{3s}\beta_K^3\left|F_K(s)\right|^2\,,
\end{eqnarray} 
obtained using the Feynman amplitude ${\cal M}^\mu_{\rm EM}$.} is described by the power-law function
\begin{eqnarray}
|F_K(s)|={2{\cal F}}/{s}    \,, \label{eq:em-ff}
\end{eqnarray}
with a constant parameter $\mathcal{F}$ predicted by the perturbative-QCD counting rule~\cite{ref_pqcd_KFF1,ref_pqcd_KFF2}, 
and consistent with previous BESIII measurement~\cite{ref_exp_KFF3}.  
From Eq.~\eqref{eq:a-tot}, the modulus of the $\psip$ strong form factor is proportional to the EM form factor through the parameter ${\cal C}/M^2$, {\it i.e.},
\begin{eqnarray}
|f_K|= g_{\psi(2S)}^\gamma\frac{\cal C}{M^2} |F_K(M^2)|
={\cal C}|F_K(M^2)|\sqrt{\frac{3\Gamma_{ee}}{\alpha M}}\,. 
    \label{eq:strong-ff}
\end{eqnarray}
To describe the data, the initial-state-radiation (ISR) and beam energy spread effect are included by convolving the radiation function $F^{\rm ISR}(x,s)$ and a Gaussian function $GS$ of zero mean and width $\sigma_E$,
\begin{eqnarray}
%\label{eq_BWG}
\sigma^{\rm obs}
%=\hspace{-3mm} \int\limits^{\sqrt{s}+n S_{\rm E}}_{\sqrt{s}-n S_{\rm E}}\hspace{-5mm} GS(\sqrt{s}-\sqrt{s^\prime}) d{\sqrt{s^\prime}} 
=\int\limits^{+\infty}_{0}GS(\sqrt{s}-\sqrt{s^\prime}) d{\sqrt{s^\prime}} 
\int \limits_{0}^{\Xf}\hspace{-2mm} F^{\rm ISR}(x,s) \sigma^{0}(s(1\!-x))dx\,, \notag
\end{eqnarray}
where $x\equiv 1-s^{\prime}/s$ and $\Xf$ is the upper limit of $x$, set at 0.11 and consistent with data.
The function $F^{\rm ISR}(x,s)$ describing ISR is defined in Eq.~(28) of Ref.~\cite{Kuraev85}. The beam energy spread parameter $\sigma_{\rm E}$ is determined as $(1.40\pm0.08)$~MeV from the fit. To improve the regression speed, instead of making a two-fold integration, the formula proposed in Ref.~\cite{ana_form} (Eq.~(22)) is used. The parameters $\mathcal{C}$, $\mathcal{F}$ and $\Phi$ are set free in the fit and the best values of them are listed in Table~\ref{tab_result}. Figure~\ref{fig_xs} shows the measured cross sections and the fit results of different solutions with individual components \footnote{The positive/negative interference contributions arise from the terms containing $e^{i\Phi}$ in the expansion of Eq.~\ref{eq_Born}.}. 
%The values of $\Phi$ in Refs.~\cite{Yuan,Metreveli,Metreveli0,ref_exp_KFF2} were all calculated under the SU(3) symmetry hypothesis with the branching fractions of $\psip\to P\bar{P}$ that neglected the continuum-resonance interference.
For the parameter $\mathcal{C}$, both solutions are consistent with the previous results under the SU(3) flavor symmetry hypothesis with improved precision.

The branching fraction of the decay $\psip\to K^+K^-$ is calculated by  
\begin{eqnarray}%\label{eq_Br}
\mathcal{B} 
=\beta^{3}_{K}\left({\mathcal{F}}/{M^2}\right)^2\left|1+\mathcal{C}e^{i\Phi}\right|^2 {\Gamma_{ee}}/{\Gamma}. \notag
\end{eqnarray}
To account for parameter correlations, the branching fraction $\mathcal{B}$ and its uncertainty are determined via multivariate Gaussian sampling, using correlation coefficients from the nominal fit. The central value of $\mathcal{B}$ is derived from its distribution (Table~\ref{tab_result}). The systematic uncertainties from the fixed parameters $M$, $\Gamma$, $\Gamma_{ee}$ are evaluated by $\pm1\sigma$ variations, while the $\Xf$-related uncertainty is estimated by varying its value. 
The largest difference for each parameter is added quadratically in the final result. 

\begin{figure}[htbp]
\begin{center}
\includegraphics[angle=0,width=.8\columnwidth]{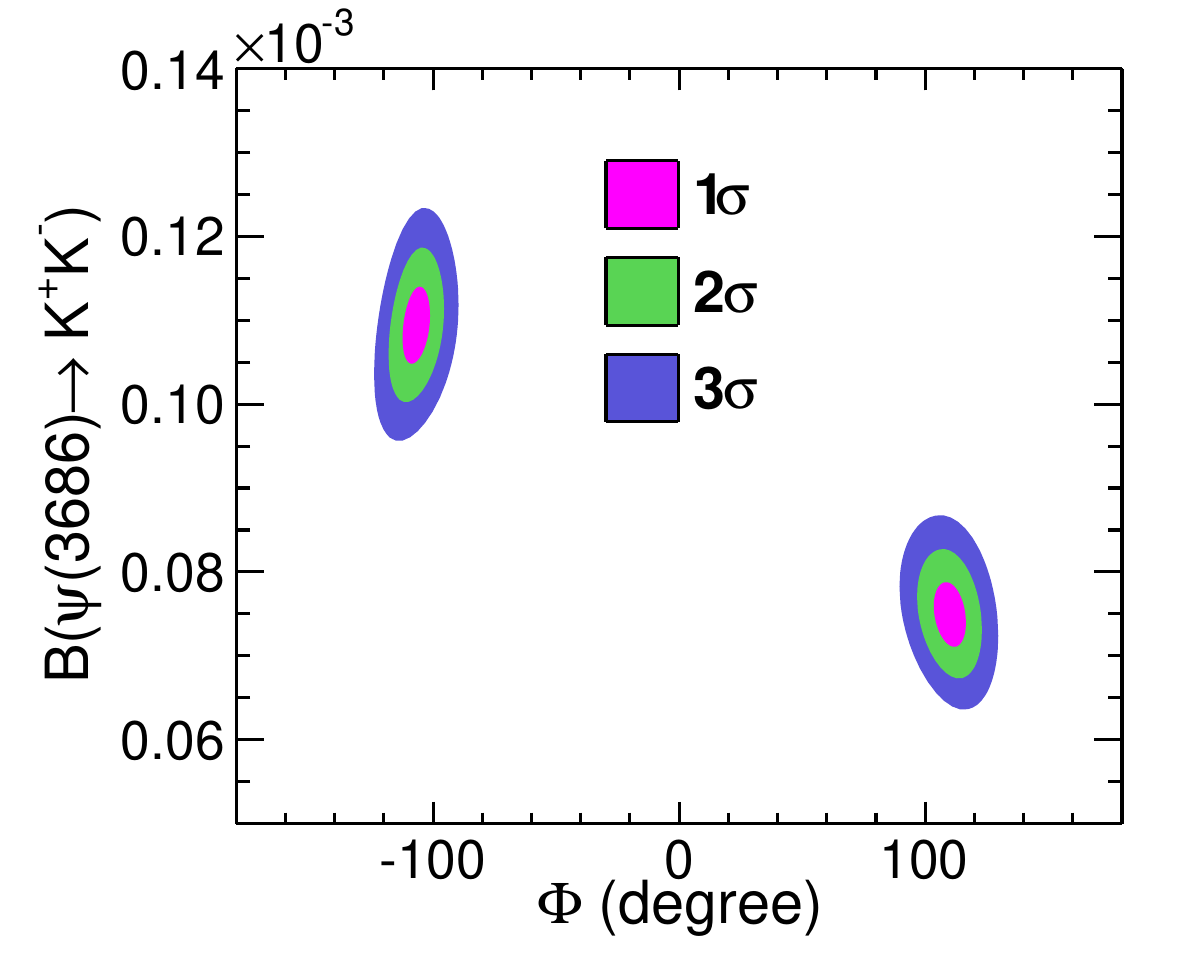} 
\caption{Contours in the $\mathcal{B}(\psip\to\kp\km)$ versus the relative phase $\Phi$ plane.}
\label{fig_contour}
\end{center}
\end{figure}

A scan in the $\mathcal{B}$ versus $\Phi$ plane is performed and the result is shown in Fig.~\ref{fig_contour}. The fit results reveal two equally-likely solutions with distinct branching fractions, whose confidence intervals do not overlap. Two physical scenarios obtained for the first time have been resolved into non-overlapping interpretations by experimental data. There is a strong correlation between $\mathcal{B}$ and $\Phi$, indeed, the branching fraction changes drastically as the phase varies.
\begin{table*}
\caption{\label{tab_result}Nominal results of the parameters from the fit of the cross section line-shape. The results are compared with the previous ones based on BES~\cite{Yuan} and CLEO data~\cite{Metreveli,Metreveli0}. The ``Seth {\it et al.}'' refers to an independent analysis using CLEO data (recalculating the $\Phi$ result with Eq.~(2) in Ref.~\cite{Metreveli}), but not affiliated with the CLEO Collaboration.}
\begin{ruledtabular}
\begin{tabular}{ccccccccc}
 \multirow{2}{*}{Parameter} &\multicolumn{2}{c}{This work} &\multirow{2}{*}{BES}  &\multirow{2}{*}{CLEO}  &\multirow{2}{*}{Seth {\it et al.}} \\
 &Positive           &Negative        \\ \hline
$\Phi$ $(^\circ)$          &$110.1 \pm6.7$      &$-106.8\pm5.7$  &$(89\pm35)^\circ$  &$(93\pm20)^\circ$  &$(66.6\pm11.5)^\circ$  \\
%$S_{\rm E}$~(MeV$/c^{2}$)  &$1.39\pm0.08$      &$1.39\pm0.08$   &$-$  &$-$  &$-$ \\
$\mathcal{C}$              &$3.18\pm0.16$      &$3.77\pm0.15$   &$2.6^{+0.9}_{-1.4}$ &$2.8^{+1.2}_{-2.8}$  &$-$   \\
$\mathcal{F}$ (GeV$^2$)    &$0.467\pm0.009$    &$0.467\pm0.009$ &$-$  &$-$  &$-$ \\
$\mathcal{B}$ ($10^{-5}$)  &$7.49\pm0.41$        &$10.94\pm0.49$    &$6.1\pm2.1$ &$6.3\pm0.7$ &$7.48\pm0.45$\\   
\end{tabular}
\end{ruledtabular}
\end{table*}

%\vspace{-5mm}

%\vspace{-5mm}

Using the definitions of Eq.~\eqref{eq:em-ff} and the ${\cal F}$ parameter value in Table~\ref{tab_result}, the modulus of the charged kaon EM form factor at the $\psip$ mass is
\begin{eqnarray}
|F_{K}(M^2)|={2\cal F}/{M^2}=0.0687\pm0.0013\,. \notag
\end{eqnarray}
Meanwhile, the strong form factor is calculated using Eq.~\eqref{eq:strong-ff} with the parameters from Table~\ref{tab_result} and the PDG values for $M$ and $\Gamma_{ee}$, yielding moduli $|f_K|_+$ and $|f_K|_-$ for the positive and negative solutions
\begin{eqnarray}
|f_{K}|_+&=& (3.53\pm0.15)\times10^{-3}\,, \notag \\
|f_{K}|_-&=& (4.18\pm0.13)\times10^{-3} \,. \notag
\label{eq:strong-results}
\end{eqnarray}

\begin{figure}[htbp]
\begin{center}
\includegraphics[angle=0,width=.8\columnwidth]{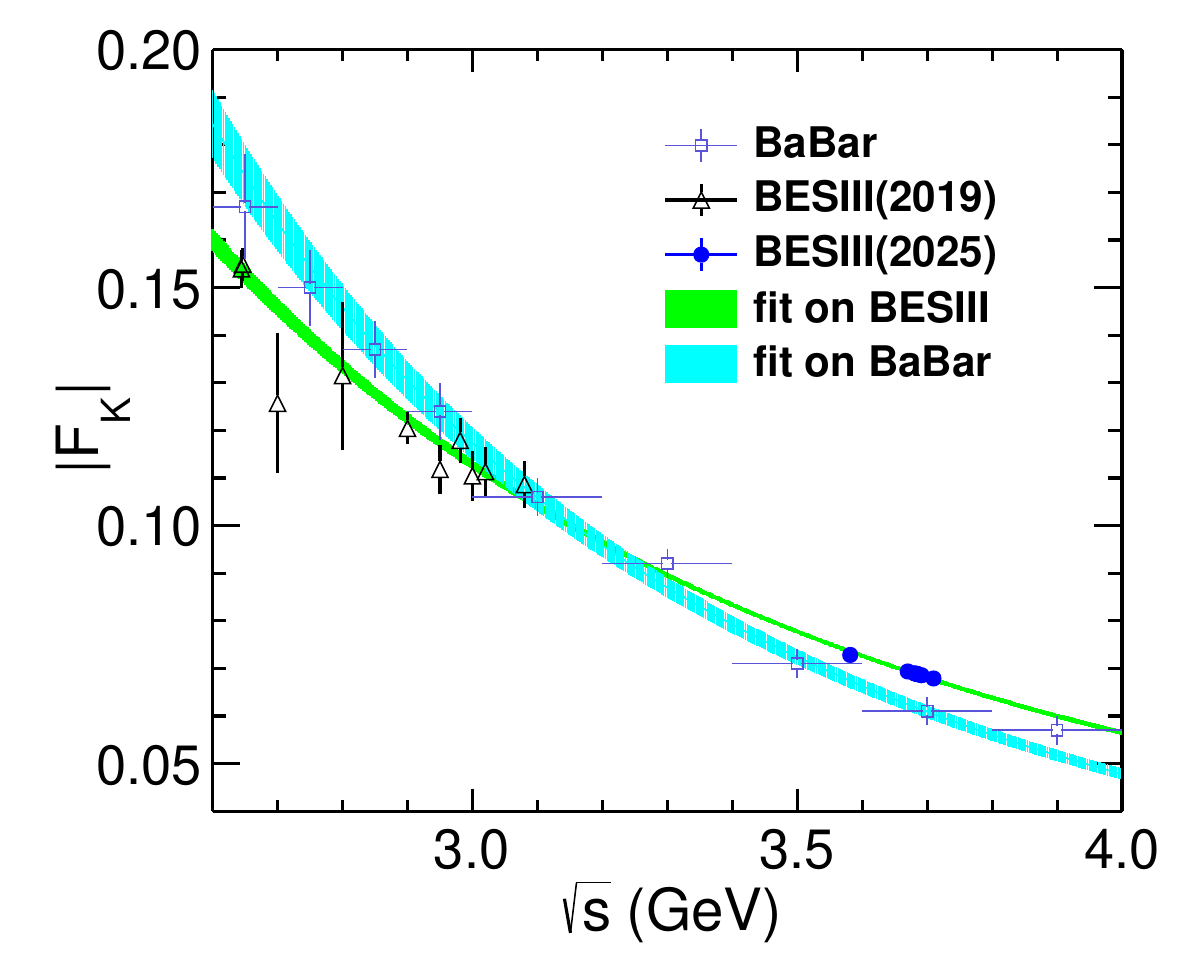}
\vspace{-5mm}\\
\caption{Comparison of our measured form factors (blue points with error bars) with the previous results from BESIII~\cite{ref_exp_KFF3} (black triangles with error bars) and BaBar~\cite{ref_exp_KFF1} (hole squares with error bars).} 
\label{fig_FF}
\end{center}
\end{figure}

The EM form factors of charged kaons are computed at all nine energy points based on the perturbative-QCD power-law behavior defined in Eq.~\eqref{eq:em-ff}. As shown in Fig.~\ref{fig_FF}, our results agree with the previous BESIII measurement~\cite{ref_exp_KFF3} and are consistent with BaBar result~\cite{ref_exp_KFF1} within $3\sigma$.
We perform a combined fit to both the BESIII data using the ansatz $|F_{K}|\propto \alpha_{s}(\sqrt{s})/s^{n}$ with $n$ free, where $\alpha_{s}(\sqrt{s})\propto 1/\ln(s/\Lambda^2)$ is the strong coupling constant with the QCD scale $\Lambda = 300$~MeV~\cite{ref_pqcd_KFF2}. The fitted parameter $n = 0.965\pm0.024$ for BESIII agrees with theoretical predictions~\cite{ref_pqcd_KFF1,ref_pqcd_KFF2} but differs significantly from that of BaBar, $n = 1.356\pm0.047$, highlighting the need for additional high-energy measurements.

%The complete set of values, including Luminosities, observed number of events, detection efficiencies, vacuum polarization factors, cross sections, and the EM form factors, are provided in the Supplementary Material.

%\vspace{-5mm}

In summary, we present the first fine scan measurement of the cross section lineshape of $\psip\to\kp\km$, with higher precision than Babar ISR~\cite{ref_exp_KFF2} and single resonance~\cite{Metreveli} data. Based on these, the relative phase between strong and EM amplitudes, the branching fraction of $\psip\to\kp\km$, the EM and the strong form factors of the charged kaon are measured. 
The relative phase between strong and EM amplitudes is determined with two distinct solutions: $\Phi_+=(110.1 \pm6.7)^{\circ}$ and $\Phi_-=(-106.8\pm5.7)^{\circ}$, corresponding to the branching fractions of $\mathcal{B}_+=(7.49\pm0.41)\times10^{-5}$ and $\mathcal{B}_-=(10.94\pm0.48)\times10^{-5}$, respectively. 
The observed correlation reveals the necessity of considering the interference effects between EM and strong amplitudes in the vicinity of the $\psip$ resonance. 
%These results present an opportunity to identify the unique physical phase through additional constraints, such as a comprehensive multi-channel analysis.
The precision of the measured $\Phi$ as well as the ratio ${\cal C}$ are improved significantly compared to the previous results. Furthermore, the modulus of the $\psip$ strong form factor of the charged kaon is also measured for the first time. 
Our measurements of the charged kaon EM form factor agree with the previous BESIII result but show systematically higher central values compared to the BaBar results, suggesting the need for more precision measurements at higher energy points.
These results shed light on long-standing ambiguities in phase extraction and provide crucial inputs for understanding charmonium decay dynamics. 

%Future precise measurements of $\psip$ parameters, particularly the branching fractions of $\psip\to\ks\kl$, will be instrumental in determining the physical solution for the phase. This work provides a foundation for further investigations into the hadronic and electromagnetic properties of charmonium states and their decay processes.

%\begin{acknowledgments}
%% Saved at => 2025-08-15
\textbf{Acknowledgement}

The BESIII Collaboration thanks the staff of BEPCII (https://cstr.cn/31109.02.BEPC) and the IHEP computing center for their strong support. This work is supported in part by National Key R\&D Program of China under Contracts Nos. 2023YFA1606000, 2023YFA1606704; Beijing Natural Science Foundation (BJNSF) under Contract No. JQ22002; National Natural Science Foundation of China (NSFC) under Contracts Nos. 11635010, 11935015, 11935016, 11935018, 12025502, 12035009, 12035013, 12061131003, 12105100, 12192260, 12192261, 12192262, 12192263, 12192264, 12192265, 12221005, 12225509, 12235017, 12361141819; the Chinese Academy of Sciences (CAS) Large-Scale Scientific Facility Program; the Strategic Priority Research Program of Chinese Academy of Sciences under Contract No. XDA0480600; CAS under Contract No. YSBR-101; 100 Talents Program of CAS; The Institute of Nuclear and Particle Physics (INPAC) and Shanghai Key Laboratory for Particle Physics and Cosmology; ERC under Contract No. 758462; German Research Foundation DFG under Contract No. FOR5327; Istituto Nazionale di Fisica Nucleare, Italy; Knut and Alice Wallenberg Foundation under Contracts Nos. 2021.0174, 2021.0299; Ministry of Development of Turkey under Contract No. DPT2006K-120470; National Research Foundation of Korea under Contract No. NRF-2022R1A2C1092335; National Science and Technology fund of Mongolia; Polish National Science Centre under Contract No. 2024/53/B/ST2/00975; STFC (United Kingdom); Swedish Research Council under Contract No. 2019.04595; U. S. Department of Energy under Contract No. DE-FG02-05ER41374

%\textbf{Other Fund Information}

%To be inserted with an additional sentence into papers that are relevant to the topic of special funding for specific topics. Authors can suggest which to Li Weiguo and/or the physics coordinator.
%        Example added sentence: This paper is also supported by the NSFC under Contract Nos. 10805053, 10979059, ....National Natural Science Foundation of China (NSFC), 10805053, PWANational Natural Science Foundation of China (NSFC), 10979059, Lund弦碎裂强子化模型及其通用强子产生器研究National Natural Science Foundation of China (NSFC), 10775075, National Natural Science Foundation of China (NSFC), 10979012, baryonsNational Natural Science Foundation of China (NSFC), 10979038, charmoniumNational Natural Science Foundation of China (NSFC), 10905034, psi(2S)->B BbarNational Natural Science Foundation of China (NSFC), 10975093, D 介子National Natural Science Foundation of China (NSFC), 10979033, psi(2S)->VPNational Natural Science Foundation of China (NSFC), 10979058, hcNational Natural Science Foundation of China (NSFC), 10975143, charmonium rare decays
%% ends here %%

%\end{acknowledgments}

\bibliography{apssamp_v10h}% Produces the bibliography via BibTeX.

\begin{thebibliography}{38}%
\makeatletter
\providecommand \@ifxundefined [1]{%
 \@ifx{#1\undefined}
}%
\providecommand \@ifnum [1]{%
 \ifnum #1\expandafter \@firstoftwo
 \else \expandafter \@secondoftwo
 \fi
}%
\providecommand \@ifx [1]{%
 \ifx #1\expandafter \@firstoftwo
 \else \expandafter \@secondoftwo
 \fi
}%
\providecommand \natexlab [1]{#1}%
\providecommand \enquote  [1]{``#1''}%
\providecommand \bibnamefont  [1]{#1}%
\providecommand \bibfnamefont [1]{#1}%
\providecommand \citenamefont [1]{#1}%
\providecommand \href@noop [0]{\@secondoftwo}%
\providecommand \href [0]{\begingroup \@sanitize@url \@href}%
\providecommand \@href[1]{\@@startlink{#1}\@@href}%
\providecommand \@@href[1]{\endgroup#1\@@endlink}%
\providecommand \@sanitize@url [0]{\catcode `\\12\catcode `\$12\catcode `\&12\catcode `\#12\catcode `\^12\catcode `\_12\catcode `\%12\relax}%
\providecommand \@@startlink[1]{}%
\providecommand \@@endlink[0]{}%
\providecommand \url  [0]{\begingroup\@sanitize@url \@url }%
\providecommand \@url [1]{\endgroup\@href {#1}{\urlprefix }}%
\providecommand \urlprefix  [0]{URL }%
\providecommand \Eprint [0]{\href }%
\providecommand \doibase [0]{http://dx.doi.org/}%
\providecommand \selectlanguage [0]{\@gobble}%
\providecommand \bibinfo  [0]{\@secondoftwo}%
\providecommand \bibfield  [0]{\@secondoftwo}%
\providecommand \translation [1]{[#1]}%
\providecommand \BibitemOpen [0]{}%
\providecommand \bibitemStop [0]{}%
\providecommand \bibitemNoStop [0]{.\EOS\space}%
\providecommand \EOS [0]{\spacefactor3000\relax}%
\providecommand \BibitemShut  [1]{\csname bibitem#1\endcsname}%
\let\auto@bib@innerbib\@empty
%</preamble>
\bibitem [{\citenamefont {Suzuki}(1998)}]{Suzuki:1998ea}%
  \BibitemOpen
  \bibfield  {author} {\bibinfo {author} {\bibfnamefont {M.}~\bibnamefont {Suzuki}},\ }\href {\doibase 10.1103/PhysRevD.57.5717} {\bibfield  {journal} {\bibinfo  {journal} {Phys. Rev. D}\ }\textbf {\bibinfo {volume} {57}},\ \bibinfo {pages} {5717} (\bibinfo {year} {1998})},\ \Eprint {http://arxiv.org/abs/hep-ph/9801284} {arXiv:hep-ph/9801284} \BibitemShut {NoStop}%
\bibitem [{\citenamefont {Suzuki}(1999)}]{Suzuki:1999nb}%
  \BibitemOpen
  \bibfield  {author} {\bibinfo {author} {\bibfnamefont {M.}~\bibnamefont {Suzuki}},\ }\href {\doibase 10.1103/PhysRevD.60.051501} {\bibfield  {journal} {\bibinfo  {journal} {Phys. Rev. D}\ }\textbf {\bibinfo {volume} {60}},\ \bibinfo {pages} {051501} (\bibinfo {year} {1999})},\ \Eprint {http://arxiv.org/abs/hep-ph/9901327} {arXiv:hep-ph/9901327} \BibitemShut {NoStop}%
\bibitem [{\citenamefont {Suzuki}(2001)}]{Suzuki:2000yq}%
  \BibitemOpen
  \bibfield  {author} {\bibinfo {author} {\bibfnamefont {M.}~\bibnamefont {Suzuki}},\ }\href {\doibase 10.1103/PhysRevD.63.054021} {\bibfield  {journal} {\bibinfo  {journal} {Phys. Rev. D}\ }\textbf {\bibinfo {volume} {63}},\ \bibinfo {pages} {054021} (\bibinfo {year} {2001})},\ \Eprint {http://arxiv.org/abs/hep-ph/0006296} {arXiv:hep-ph/0006296} \BibitemShut {NoStop}%
\bibitem [{\citenamefont {Rosner}(1999)}]{Rosner}%
  \BibitemOpen
  \bibfield  {author} {\bibinfo {author} {\bibfnamefont {J.~L.}\ \bibnamefont {Rosner}},\ }\href {\doibase 10.1103/PhysRevD.60.074029} {\bibfield  {journal} {\bibinfo  {journal} {Phys. Rev. D}\ }\textbf {\bibinfo {volume} {60}},\ \bibinfo {pages} {074029} (\bibinfo {year} {1999})},\ \Eprint {http://arxiv.org/abs/hep-ph/9903543} {arXiv:hep-ph/9903543} \BibitemShut {NoStop}%
\bibitem [{\citenamefont {Baldini}\ \emph {et~al.}(1998)\citenamefont {Baldini} \emph {et~al.}}]{Baldini:1998en}%
  \BibitemOpen
  \bibfield  {author} {\bibinfo {author} {\bibfnamefont {R.}~\bibnamefont {Baldini}} \emph {et~al.},\ }\href {\doibase 10.1016/S0370-2693(98)01358-6} {\bibfield  {journal} {\bibinfo  {journal} {Phys. Lett. B}\ }\textbf {\bibinfo {volume} {444}},\ \bibinfo {pages} {111} (\bibinfo {year} {1998})}\BibitemShut {NoStop}%
\bibitem [{\citenamefont {Baldini}\ \emph {et~al.}(1997)\citenamefont {Baldini}, \citenamefont {Bini},\ and\ \citenamefont {Luppi}}]{Baldini:1996hc}%
  \BibitemOpen
  \bibfield  {author} {\bibinfo {author} {\bibfnamefont {R.}~\bibnamefont {Baldini}}, \bibinfo {author} {\bibfnamefont {C.}~\bibnamefont {Bini}}, \ and\ \bibinfo {author} {\bibfnamefont {E.}~\bibnamefont {Luppi}},\ }\href {\doibase 10.1016/S0370-2693(97)00614-X} {\bibfield  {journal} {\bibinfo  {journal} {Phys. Lett. B}\ }\textbf {\bibinfo {volume} {404}},\ \bibinfo {pages} {362} (\bibinfo {year} {1997})}\BibitemShut {NoStop}%
\bibitem [{\citenamefont {Gerard}\ and\ \citenamefont {Weyers}(1999)}]{Gerard}%
  \BibitemOpen
  \bibfield  {author} {\bibinfo {author} {\bibfnamefont {J.~M.}\ \bibnamefont {Gerard}}\ and\ \bibinfo {author} {\bibfnamefont {J.}~\bibnamefont {Weyers}},\ }\href {\doibase 10.1016/S0370-2693(99)00849-7} {\bibfield  {journal} {\bibinfo  {journal} {Phys. Lett. B}\ }\textbf {\bibinfo {volume} {462}},\ \bibinfo {pages} {324} (\bibinfo {year} {1999})},\ \Eprint {http://arxiv.org/abs/hep-ph/9906357} {arXiv:hep-ph/9906357} \BibitemShut {NoStop}%
\bibitem [{\citenamefont {Haber}\ and\ \citenamefont {Perrier}(1985)}]{Haber:1985cv}%
  \BibitemOpen
  \bibfield  {author} {\bibinfo {author} {\bibfnamefont {H.~E.}\ \bibnamefont {Haber}}\ and\ \bibinfo {author} {\bibfnamefont {J.}~\bibnamefont {Perrier}},\ }\href {\doibase 10.1103/PhysRevD.32.2961} {\bibfield  {journal} {\bibinfo  {journal} {Phys. Rev. D}\ }\textbf {\bibinfo {volume} {32}},\ \bibinfo {pages} {2961} (\bibinfo {year} {1985})}\BibitemShut {NoStop}%
\bibitem [{\citenamefont {Metreveli}\ \emph {et~al.}(2012)\citenamefont {Metreveli}, \citenamefont {Dobbs}, \citenamefont {Tomaradze}, \citenamefont {Xiao}, \citenamefont {Seth}, \citenamefont {Yelton}, \citenamefont {Asner}, \citenamefont {Tatishvili},\ and\ \citenamefont {Bonvicini}}]{Metreveli}%
  \BibitemOpen
  \bibfield  {author} {\bibinfo {author} {\bibfnamefont {Z.}~\bibnamefont {Metreveli}}, \bibinfo {author} {\bibfnamefont {S.}~\bibnamefont {Dobbs}}, \bibinfo {author} {\bibfnamefont {A.}~\bibnamefont {Tomaradze}}, \bibinfo {author} {\bibfnamefont {T.}~\bibnamefont {Xiao}}, \bibinfo {author} {\bibfnamefont {K.~K.}\ \bibnamefont {Seth}}, \bibinfo {author} {\bibfnamefont {J.}~\bibnamefont {Yelton}}, \bibinfo {author} {\bibfnamefont {D.~M.}\ \bibnamefont {Asner}}, \bibinfo {author} {\bibfnamefont {G.}~\bibnamefont {Tatishvili}}, \ and\ \bibinfo {author} {\bibfnamefont {G.}~\bibnamefont {Bonvicini}},\ }\href {\doibase 10.1103/PhysRevD.85.092007} {\bibfield  {journal} {\bibinfo  {journal} {Phys. Rev. D}\ }\textbf {\bibinfo {volume} {85}},\ \bibinfo {pages} {092007} (\bibinfo {year} {2012})},\ \Eprint {http://arxiv.org/abs/1203.5361} {arXiv:1203.5361 [hep-ex]} \BibitemShut {NoStop}%
\bibitem [{\citenamefont {Li}(2008)}]{Li:2008fk}%
  \BibitemOpen
  \bibfield  {author} {\bibinfo {author} {\bibfnamefont {X.-Q.}\ \bibnamefont {Li}},\ }\href@noop {} {\bibfield  {journal} {\bibinfo  {journal} {arXiv:0812.5037 [hep-ph]}\ } (\bibinfo {year} {2008})},\ \Eprint {http://arxiv.org/abs/0812.5037} {arXiv:0812.5037 [hep-ph]} \BibitemShut {NoStop}%
\bibitem [{\citenamefont {Zhao}\ \emph {et~al.}(2010)\citenamefont {Zhao}, \citenamefont {Li},\ and\ \citenamefont {Chang}}]{Zhao:2010zzv}%
  \BibitemOpen
  \bibfield  {author} {\bibinfo {author} {\bibfnamefont {Q.}~\bibnamefont {Zhao}}, \bibinfo {author} {\bibfnamefont {G.}~\bibnamefont {Li}}, \ and\ \bibinfo {author} {\bibfnamefont {C.-H.}\ \bibnamefont {Chang}},\ }\href {\doibase 10.1088/1674-1137/34/2/027} {\bibfield  {journal} {\bibinfo  {journal} {Chin. Phys. C}\ }\textbf {\bibinfo {volume} {34}},\ \bibinfo {pages} {299} (\bibinfo {year} {2010})}\BibitemShut {NoStop}%
\bibitem [{\citenamefont {Kivel}(2023)}]{Kivel:2023fgu}%
  \BibitemOpen
  \bibfield  {author} {\bibinfo {author} {\bibfnamefont {N.}~\bibnamefont {Kivel}},\ }\href {\doibase 10.1103/PhysRevD.107.094015} {\bibfield  {journal} {\bibinfo  {journal} {Phys. Rev. D}\ }\textbf {\bibinfo {volume} {107}},\ \bibinfo {pages} {094015} (\bibinfo {year} {2023})},\ \Eprint {http://arxiv.org/abs/2301.03884} {arXiv:2301.03884 [hep-ph]} \BibitemShut {NoStop}%
\bibitem [{\citenamefont {Wang}\ \emph {et~al.}(2004)\citenamefont {Wang}, \citenamefont {Yuan}, \citenamefont {Mo},\ and\ \citenamefont {Zhang}}]{Wang:2002np}%
  \BibitemOpen
  \bibfield  {author} {\bibinfo {author} {\bibfnamefont {P.}~\bibnamefont {Wang}}, \bibinfo {author} {\bibfnamefont {C.~Z.}\ \bibnamefont {Yuan}}, \bibinfo {author} {\bibfnamefont {X.~H.}\ \bibnamefont {Mo}}, \ and\ \bibinfo {author} {\bibfnamefont {D.~H.}\ \bibnamefont {Zhang}},\ }\href {\doibase 10.1016/j.physletb.2004.04.073} {\bibfield  {journal} {\bibinfo  {journal} {Phys. Lett. B}\ }\textbf {\bibinfo {volume} {593}},\ \bibinfo {pages} {89} (\bibinfo {year} {2004})},\ \Eprint {http://arxiv.org/abs/hep-ex/0210063} {arXiv:hep-ex/0210063} \BibitemShut {NoStop}%
\bibitem [{\citenamefont {L{\'o}pez~Castro}\ \emph {et~al.}(1995)\citenamefont {L{\'o}pez~Castro}, \citenamefont {Lucio~M.},\ and\ \citenamefont {Pestieau}}]{LopezCastro:1994xw}%
  \BibitemOpen
  \bibfield  {author} {\bibinfo {author} {\bibfnamefont {G.}~\bibnamefont {L{\'o}pez~Castro}}, \bibinfo {author} {\bibfnamefont {J.~L.}\ \bibnamefont {Lucio~M.}}, \ and\ \bibinfo {author} {\bibfnamefont {J.}~\bibnamefont {Pestieau}},\ }\href {\doibase 10.1063/1.48783} {\bibfield  {journal} {\bibinfo  {journal} {AIP Conf. Proc.}\ }\textbf {\bibinfo {volume} {342}},\ \bibinfo {pages} {441} (\bibinfo {year} {1995})},\ \Eprint {http://arxiv.org/abs/hep-ph/9902300} {arXiv:hep-ph/9902300} \BibitemShut {NoStop}%
\bibitem [{\citenamefont {Appelquist}\ and\ \citenamefont {Politzer}(1975)}]{Appelquist}%
  \BibitemOpen
  \bibfield  {author} {\bibinfo {author} {\bibfnamefont {T.}~\bibnamefont {Appelquist}}\ and\ \bibinfo {author} {\bibfnamefont {H.~D.}\ \bibnamefont {Politzer}},\ }\href {\doibase 10.1103/PhysRevLett.34.43} {\bibfield  {journal} {\bibinfo  {journal} {Phys. Rev. Lett.}\ }\textbf {\bibinfo {volume} {34}},\ \bibinfo {pages} {43} (\bibinfo {year} {1975})}\BibitemShut {NoStop}%
\bibitem [{\citenamefont {Yuan}\ \emph {et~al.}(2003)\citenamefont {Yuan}, \citenamefont {Wang},\ and\ \citenamefont {Mo}}]{Yuan}%
  \BibitemOpen
  \bibfield  {author} {\bibinfo {author} {\bibfnamefont {C.~Z.}\ \bibnamefont {Yuan}}, \bibinfo {author} {\bibfnamefont {P.}~\bibnamefont {Wang}}, \ and\ \bibinfo {author} {\bibfnamefont {X.~H.}\ \bibnamefont {Mo}},\ }\href {\doibase 10.1016/j.physletb.2003.06.023} {\bibfield  {journal} {\bibinfo  {journal} {Phys. Lett. B}\ }\textbf {\bibinfo {volume} {567}},\ \bibinfo {pages} {73} (\bibinfo {year} {2003})},\ \Eprint {http://arxiv.org/abs/hep-ph/0305259} {arXiv:hep-ph/0305259} \BibitemShut {NoStop}%
\bibitem [{\citenamefont {Bai}\ \emph {et~al.}(2004)\citenamefont {Bai} \emph {et~al.}}]{Bai}%
  \BibitemOpen
  \bibfield  {author} {\bibinfo {author} {\bibfnamefont {J.~Z.}\ \bibnamefont {Bai}} \emph {et~al.} (\bibinfo {collaboration} {BES}),\ }\href {\doibase 10.1103/PhysRevLett.92.052001} {\bibfield  {journal} {\bibinfo  {journal} {Phys. Rev. Lett.}\ }\textbf {\bibinfo {volume} {92}},\ \bibinfo {pages} {052001} (\bibinfo {year} {2004})},\ \Eprint {http://arxiv.org/abs/hep-ex/0310024} {arXiv:hep-ex/0310024} \BibitemShut {NoStop}%
\bibitem [{\citenamefont {Dobbs}\ \emph {et~al.}(2006)\citenamefont {Dobbs} \emph {et~al.}}]{Metreveli0}%
  \BibitemOpen
  \bibfield  {author} {\bibinfo {author} {\bibfnamefont {S.}~\bibnamefont {Dobbs}} \emph {et~al.} (\bibinfo {collaboration} {CLEO}),\ }\href {\doibase 10.1103/PhysRevD.74.011105} {\bibfield  {journal} {\bibinfo  {journal} {Phys. Rev. D}\ }\textbf {\bibinfo {volume} {74}},\ \bibinfo {pages} {011105} (\bibinfo {year} {2006})},\ \Eprint {http://arxiv.org/abs/hep-ex/0603020} {arXiv:hep-ex/0603020} \BibitemShut {NoStop}%
\bibitem [{\citenamefont {Lees}\ \emph {et~al.}(2015)\citenamefont {Lees} \emph {et~al.}}]{ref_exp_KFF2}%
  \BibitemOpen
  \bibfield  {author} {\bibinfo {author} {\bibfnamefont {J.~P.}\ \bibnamefont {Lees}} \emph {et~al.} (\bibinfo {collaboration} {BaBar}),\ }\href {\doibase 10.1103/PhysRevD.92.072008} {\bibfield  {journal} {\bibinfo  {journal} {Phys. Rev. D}\ }\textbf {\bibinfo {volume} {92}},\ \bibinfo {pages} {072008} (\bibinfo {year} {2015})},\ \Eprint {http://arxiv.org/abs/1507.04638} {arXiv:1507.04638 [hep-ex]} \BibitemShut {NoStop}%
\bibitem [{\citenamefont {Anashin}\ \emph {et~al.}(2010)\citenamefont {Anashin} \emph {et~al.}}]{KEDR}%
  \BibitemOpen
  \bibfield  {author} {\bibinfo {author} {\bibfnamefont {V.~V.}\ \bibnamefont {Anashin}} \emph {et~al.} (\bibinfo {collaboration} {KEDR}),\ }\href {\doibase 10.1016/j.physletb.2010.01.057} {\bibfield  {journal} {\bibinfo  {journal} {Phys. Lett. B}\ }\textbf {\bibinfo {volume} {685}},\ \bibinfo {pages} {134} (\bibinfo {year} {2010})},\ \Eprint {http://arxiv.org/abs/0912.1082} {arXiv:0912.1082 [hep-ex]} \BibitemShut {NoStop}%
\bibitem [{\citenamefont {Ablikim}\ \emph {et~al.}(2019{\natexlab{a}})\citenamefont {Ablikim} \emph {et~al.}}]{J5pi}%
  \BibitemOpen
  \bibfield  {author} {\bibinfo {author} {\bibfnamefont {M.}~\bibnamefont {Ablikim}} \emph {et~al.} (\bibinfo {collaboration} {BESIII}),\ }\href {\doibase 10.1016/j.physletb.2019.03.001} {\bibfield  {journal} {\bibinfo  {journal} {Phys. Lett. B}\ }\textbf {\bibinfo {volume} {791}},\ \bibinfo {pages} {375} (\bibinfo {year} {2019}{\natexlab{a}})},\ \Eprint {http://arxiv.org/abs/1808.02166} {arXiv:1808.02166 [hep-ex]} \BibitemShut {NoStop}%
\bibitem [{\citenamefont {Rosini}\ and\ \citenamefont {Pacetti}(2025)}]{Rosini:2025wfm}%
  \BibitemOpen
  \bibfield  {author} {\bibinfo {author} {\bibfnamefont {F.}~\bibnamefont {Rosini}}\ and\ \bibinfo {author} {\bibfnamefont {S.}~\bibnamefont {Pacetti}},\ }\href {\doibase 10.1140/epjc/s10052-025-13976-7} {\bibfield  {journal} {\bibinfo  {journal} {Eur. Phys. J. C}\ }\textbf {\bibinfo {volume} {85}},\ \bibinfo {pages} {236} (\bibinfo {year} {2025})},\ \Eprint {http://arxiv.org/abs/2501.04449} {arXiv:2501.04449 [hep-ph]} \BibitemShut {NoStop}%
\bibitem [{\citenamefont {Ablikim}\ \emph {et~al.}(2017)\citenamefont {Ablikim} \emph {et~al.}}]{BESIII:2017lkp}%
  \BibitemOpen
  \bibfield  {author} {\bibinfo {author} {\bibfnamefont {M.}~\bibnamefont {Ablikim}} \emph {et~al.} (\bibinfo {collaboration} {BESIII}),\ }\href {\doibase 10.1088/1674-1137/41/6/063001} {\bibfield  {journal} {\bibinfo  {journal} {Chin. Phys. C}\ }\textbf {\bibinfo {volume} {41}},\ \bibinfo {pages} {063001} (\bibinfo {year} {2017})},\ \Eprint {http://arxiv.org/abs/1702.04977} {arXiv:1702.04977 [hep-ex]} \BibitemShut {NoStop}%
\bibitem [{\citenamefont {Ablikim}\ \emph {et~al.}(2010)\citenamefont {Ablikim} \emph {et~al.}}]{BESIII}%
  \BibitemOpen
  \bibfield  {author} {\bibinfo {author} {\bibfnamefont {M.}~\bibnamefont {Ablikim}} \emph {et~al.} (\bibinfo {collaboration} {BESIII}),\ }\href {\doibase 10.1016/j.nima.2009.12.050} {\bibfield  {journal} {\bibinfo  {journal} {Nucl. Instrum. Meth. A}\ }\textbf {\bibinfo {volume} {614}},\ \bibinfo {pages} {345} (\bibinfo {year} {2010})},\ \Eprint {http://arxiv.org/abs/0911.4960} {arXiv:0911.4960 [physics.ins-det]} \BibitemShut {NoStop}%
\bibitem [{\citenamefont {Ping}(2014)}]{conexc}%
  \BibitemOpen
  \bibfield  {author} {\bibinfo {author} {\bibfnamefont {R.~G.}\ \bibnamefont {Ping}},\ }\href {\doibase 10.1088/1674-1137/38/8/083001} {\bibfield  {journal} {\bibinfo  {journal} {Chin. Phys. C}\ }\textbf {\bibinfo {volume} {38}},\ \bibinfo {pages} {083001} (\bibinfo {year} {2014})},\ \Eprint {http://arxiv.org/abs/1309.3932} {arXiv:1309.3932 [hep-ph]} \BibitemShut {NoStop}%
\bibitem [{\citenamefont {Carloni~Calame}\ \emph {et~al.}(2019)\citenamefont {Carloni~Calame}, \citenamefont {Montagna}, \citenamefont {Nicrosini},\ and\ \citenamefont {Piccinini}}]{babayaga}%
  \BibitemOpen
  \bibfield  {author} {\bibinfo {author} {\bibfnamefont {C.~M.}\ \bibnamefont {Carloni~Calame}}, \bibinfo {author} {\bibfnamefont {G.}~\bibnamefont {Montagna}}, \bibinfo {author} {\bibfnamefont {O.}~\bibnamefont {Nicrosini}}, \ and\ \bibinfo {author} {\bibfnamefont {F.}~\bibnamefont {Piccinini}},\ }\href {\doibase 10.1051/epjconf/201921807004} {\bibfield  {journal} {\bibinfo  {journal} {EPJ Web Conf.}\ }\textbf {\bibinfo {volume} {218}},\ \bibinfo {pages} {07004} (\bibinfo {year} {2019})}\BibitemShut {NoStop}%
\bibitem [{\citenamefont {Ablikim}\ \emph {et~al.}(2022)\citenamefont {Ablikim} \emph {et~al.}}]{lund}%
  \BibitemOpen
  \bibfield  {author} {\bibinfo {author} {\bibfnamefont {M.}~\bibnamefont {Ablikim}} \emph {et~al.} (\bibinfo {collaboration} {BESIII}),\ }\href {\doibase 10.1103/PhysRevLett.128.062004} {\bibfield  {journal} {\bibinfo  {journal} {Phys. Rev. Lett.}\ }\textbf {\bibinfo {volume} {128}},\ \bibinfo {pages} {062004} (\bibinfo {year} {2022})},\ \Eprint {http://arxiv.org/abs/2112.11728} {arXiv:2112.11728 [hep-ex]} \BibitemShut {NoStop}%
\bibitem [{\citenamefont {Agostinelli}\ \emph {et~al.}(2003)\citenamefont {Agostinelli} \emph {et~al.}}]{Geant4}%
  \BibitemOpen
  \bibfield  {author} {\bibinfo {author} {\bibfnamefont {S.}~\bibnamefont {Agostinelli}} \emph {et~al.} (\bibinfo {collaboration} {GEANT4}),\ }\href {\doibase 10.1016/S0168-9002(03)01368-8} {\bibfield  {journal} {\bibinfo  {journal} {Nucl. Instrum. Meth. A}\ }\textbf {\bibinfo {volume} {506}},\ \bibinfo {pages} {250} (\bibinfo {year} {2003})}\BibitemShut {NoStop}%
\bibitem [{\citenamefont {Wang}\ \emph {et~al.}(2024)\citenamefont {Wang}, \citenamefont {Wang},\ and\ \citenamefont {Wang}}]{ana_form}%
  \BibitemOpen
  \bibfield  {author} {\bibinfo {author} {\bibfnamefont {Y.-N.}\ \bibnamefont {Wang}}, \bibinfo {author} {\bibfnamefont {Y.-D.}\ \bibnamefont {Wang}}, \ and\ \bibinfo {author} {\bibfnamefont {P.}~\bibnamefont {Wang}},\ }\href {\doibase 10.1088/1674-1137/ad6c09} {\bibfield  {journal} {\bibinfo  {journal} {Chin. Phys. C}\ }\textbf {\bibinfo {volume} {48}},\ \bibinfo {pages} {113104} (\bibinfo {year} {2024})},\ \bibinfo {note} {[Erratum: Chin.Phys.C 49, 039001 (2025)]},\ \Eprint {http://arxiv.org/abs/2311.13292} {arXiv:2311.13292 [hep-ph]} \BibitemShut {NoStop}%
\bibitem [{\citenamefont {Navas}\ \emph {et~al.}(2024)\citenamefont {Navas} \emph {et~al.}}]{pdg}%
  \BibitemOpen
  \bibfield  {author} {\bibinfo {author} {\bibfnamefont {S.}~\bibnamefont {Navas}} \emph {et~al.} (\bibinfo {collaboration} {Particle Data Group}),\ }\href {\doibase 10.1103/PhysRevD.110.030001} {\bibfield  {journal} {\bibinfo  {journal} {Phys. Rev. D}\ }\textbf {\bibinfo {volume} {110}},\ \bibinfo {pages} {030001} (\bibinfo {year} {2024})}\BibitemShut {NoStop}%
\bibitem [{\citenamefont {Berends}\ and\ \citenamefont {Komen}(1976)}]{Berends}%
  \BibitemOpen
  \bibfield  {author} {\bibinfo {author} {\bibfnamefont {F.~A.}\ \bibnamefont {Berends}}\ and\ \bibinfo {author} {\bibfnamefont {G.~J.}\ \bibnamefont {Komen}},\ }\href {\doibase 10.1016/0370-2693(76)90389-0} {\bibfield  {journal} {\bibinfo  {journal} {Phys. Lett. B}\ }\textbf {\bibinfo {volume} {63}},\ \bibinfo {pages} {432} (\bibinfo {year} {1976})}\BibitemShut {NoStop}%
\bibitem [{Note1()}]{Note1}%
  \BibitemOpen
  \bibinfo {note} {The factor of two comes from the comparison between the Born cross section of Eq.~\protect \eqref {eq_Born} and \begin {eqnarray} \sigma ^0(\protect \sqrt {s})=\protect \frac {\pi \alpha ^2}{3s}\beta _K^3\left |F_K(s)\right |^2\protect \,, \end {eqnarray} obtained using the Feynman amplitude ${\protect \cal M}^\mu _{\protect \rm EM}$.}\BibitemShut {Stop}%
\bibitem [{\citenamefont {Chernyak}\ \emph {et~al.}(1977)\citenamefont {Chernyak}, \citenamefont {Zhitnitsky},\ and\ \citenamefont {Serbo}}]{ref_pqcd_KFF1}%
  \BibitemOpen
  \bibfield  {author} {\bibinfo {author} {\bibfnamefont {V.~L.}\ \bibnamefont {Chernyak}}, \bibinfo {author} {\bibfnamefont {A.~R.}\ \bibnamefont {Zhitnitsky}}, \ and\ \bibinfo {author} {\bibfnamefont {V.~G.}\ \bibnamefont {Serbo}},\ }\href@noop {} {\bibfield  {journal} {\bibinfo  {journal} {JETP Lett.}\ }\textbf {\bibinfo {volume} {26}},\ \bibinfo {pages} {594} (\bibinfo {year} {1977})}\BibitemShut {NoStop}%
\bibitem [{\citenamefont {Lepage}\ and\ \citenamefont {Brodsky}(1979)}]{ref_pqcd_KFF2}%
  \BibitemOpen
  \bibfield  {author} {\bibinfo {author} {\bibfnamefont {G.~P.}\ \bibnamefont {Lepage}}\ and\ \bibinfo {author} {\bibfnamefont {S.~J.}\ \bibnamefont {Brodsky}},\ }\href {\doibase 10.1016/0370-2693(79)90554-9} {\bibfield  {journal} {\bibinfo  {journal} {Phys. Lett. B}\ }\textbf {\bibinfo {volume} {87}},\ \bibinfo {pages} {359} (\bibinfo {year} {1979})}\BibitemShut {NoStop}%
\bibitem [{\citenamefont {Ablikim}\ \emph {et~al.}(2019{\natexlab{b}})\citenamefont {Ablikim} \emph {et~al.}}]{ref_exp_KFF3}%
  \BibitemOpen
  \bibfield  {author} {\bibinfo {author} {\bibfnamefont {M.}~\bibnamefont {Ablikim}} \emph {et~al.} (\bibinfo {collaboration} {BESIII}),\ }\href {\doibase 10.1103/PhysRevD.99.032001} {\bibfield  {journal} {\bibinfo  {journal} {Phys. Rev. D}\ }\textbf {\bibinfo {volume} {99}},\ \bibinfo {pages} {032001} (\bibinfo {year} {2019}{\natexlab{b}})},\ \Eprint {http://arxiv.org/abs/1811.08742} {arXiv:1811.08742 [hep-ex]} \BibitemShut {NoStop}%
\bibitem [{\citenamefont {Kuraev}\ and\ \citenamefont {Fadin}(1985)}]{Kuraev85}%
  \BibitemOpen
  \bibfield  {author} {\bibinfo {author} {\bibfnamefont {E.~A.}\ \bibnamefont {Kuraev}}\ and\ \bibinfo {author} {\bibfnamefont {V.~S.}\ \bibnamefont {Fadin}},\ }\href@noop {} {\bibfield  {journal} {\bibinfo  {journal} {Sov. J. Nucl. Phys.}\ }\textbf {\bibinfo {volume} {41}},\ \bibinfo {pages} {466} (\bibinfo {year} {1985})}\BibitemShut {NoStop}%
\bibitem [{Note2()}]{Note2}%
  \BibitemOpen
  \bibinfo {note} {The positive/negative interference contributions arise from the terms containing $e^{i\Phi }$ in the expansion of Eq.~\ref {eq_Born}.}\BibitemShut {Stop}%
\bibitem [{\citenamefont {Lees}\ \emph {et~al.}(2013)\citenamefont {Lees} \emph {et~al.}}]{ref_exp_KFF1}%
  \BibitemOpen
  \bibfield  {author} {\bibinfo {author} {\bibfnamefont {J.~P.}\ \bibnamefont {Lees}} \emph {et~al.} (\bibinfo {collaboration} {BaBar}),\ }\href {\doibase 10.1103/PhysRevD.88.032013} {\bibfield  {journal} {\bibinfo  {journal} {Phys. Rev. D}\ }\textbf {\bibinfo {volume} {88}},\ \bibinfo {pages} {032013} (\bibinfo {year} {2013})},\ \Eprint {http://arxiv.org/abs/1306.3600} {arXiv:1306.3600 [hep-ex]} \BibitemShut {NoStop}%
\end{thebibliography}%

\end{document}